\documentclass[preprint,11pt,aps,draft,nofootinbib,showpacs,showkeys]{revtex4}
\usepackage{amssymb}

\begin{document}

\title{Macroscopic time evolution and MaxEnt inference for closed systems with Hamiltonian dynamics}

\author{Domagoj Kui\'{c}}
\email{dkuic@pmfst.hr} 
\author{Pa\v{s}ko \v{Z}upanovi\'{c}}
\email{pasko@pmfst.hr}
\author{Davor Jureti\'{c}}
\email{juretic@pmfst.hr}
\affiliation{University of Split, Faculty of Science, N. Tesle 12, 21000 Split, Croatia}

\begin{abstract}
MaxEnt inference algorithm and information theory are relevant for the time evolution of macroscopic systems considered as problem of incomplete information. Two different MaxEnt approaches are introduced in this work, both applied to prediction of time evolution for closed Hamiltonian systems. The first one is based on Liouville equation for the conditional probability distribution, introduced as a strict microscopic constraint on time evolution in phase space. The conditional probability distribution is defined for the set of microstates associated with the set of phase space paths determined by solutions of Hamilton's equations. The MaxEnt inference algorithm with Shannon's concept of the conditional information entropy is then applied to prediction, consistently with this strict microscopic constraint on time evolution in phase space. The second approach is based on the same concepts, with a difference that Liouville equation for the conditional probability distribution is introduced as a macroscopic constraint given by a phase space average. We consider the incomplete nature of our information about microscopic dynamics in a rational way that is consistent with Jaynes' formulation of predictive statistical mechanics, and the concept of macroscopic reproducibility for time dependent processes. Maximization of the conditional information entropy subject to this macroscopic constraint leads to a loss of correlation between the initial phase space paths and final microstates. Information entropy is the theoretic upper bound on the conditional information entropy, with the upper bound attained only in case of the complete loss of correlation. In this alternative approach to prediction of macroscopic time evolution, maximization of the conditional information entropy is equivalent to the loss of statistical correlation, and leads to corresponding loss of information. In accordance with the original idea of Jaynes, irreversibility appears as a consequence of gradual loss of information about possible microstates of the system.
\end{abstract}

\pacs{02.50.Tt, 05.20.Gg, 05.70.Ln, 45.20.Jj}
\keywords{MaxEnt algorithm, information theory, statistical mechanics, Hamiltonian dynamics}

\maketitle

\section{Introduction} 

Maximum-entropy formalism, or alternatively MaxEnt algorithm, was formulated by E. T. Jaynes in his influential papers \cite{jaynes1,jaynes2} intended for applications in statistical mechanics. In Jaynes' approach a full development of the results of equilibrium statistical mechanics and formalism of Gibbs \cite{gibbs} was possible as a form of statistical inference based on Shannon's concept of information-theory entropy as a measure of information \cite{shannon}. In the language of Jaynes, it is the correct measure of the ``amount of uncertainty" in the probability distribution \cite{jaynes3}. Maximization of information-theory entropy subject to certain constraints is a central concept in Jaynes' approach, and provides the least biased probability estimates subject to the available information. It is important that Jaynes sees Gibbs' formalism as essential tool for statistical inference in different problems with insufficient information. This includes equilibrium statistical mechanics \cite{jaynes1} and the formulation of a theory of irreversibility \cite{jaynes2}, that Jaynes tries to accomplish in his later works \cite{jaynes3,jaynes4,jaynes5,jaynes6,jaynes7}.         

Predictions and calculations for different irreversible processes usually involve three distinct stages \cite{jaynes5}: (1) Setting up an ``ensemble", i.e., choosing an initial density matrix, or in our case an $N$-particle distribution function, which is to describe our initial knowledge about the system of interest; (2) Solving the dynamical problem; i.e., applying the microscopic equations of motion to obtain the time evolution of the system; (3) Extracting the final physical predictions from the time developed ensemble.  As fully recognized by Jaynes, the stage (1) and the availability of its general solution simplifies the complicated stage (2). The problem includes also an equally important stage (0) consisting of some kind of measurement or observation defining both the system and the problem \cite{grandy1}. In direct mathematical attempts that lead to a theory of irreversibility, the Liouville theorem with the conservation of phase space volume inherent to Hamiltonian dynamics, is represented as one of the main difficulties. Relation of the Liouville equation and irreversible macroscopic behavior is one of the central problems in statistical mechanics. For this reason it is reduced to an irreversible equation termed Boltzmann equation, rate equation or master equation. Far from creating difficulties, Jaynes considers the Liouville equation and the related constancy in time of Gibbs' entropy as precisely the dynamical property needed for solution of this problem, considering it to be more of conceptual than mathematical nature \cite{jaynes4}. 

Mathematical clarity of this viewpoint has its basis in a limit theorem noted by Shannon \cite{shannon}, an application of the fundamental asymptotic equipartition theorem of information theory. This theorem relates the Boltzmann's original formula for entropy of a macrostate and the Gibbs expression for entropy in the limit of a large number of particles \cite{jaynes4,jaynes5,jaynes7}. Mathematical connection with the Boltzmann's interpretation of entropy as the logarithm of the number or ways (or microstates) by which a macroscopic state can be realized, $S = k\log W$, introduces then a simple physical interpretation to the Gibbs' formalism, and its generalizations in the maximum-entropy formalism.  Maximization of the information entropy subject to constraints then predicts the macroscopic behavior that can happen in the greatest number of ways compatible with the available information. In application to time dependent processes, this is referred to by Jaynes as the maximum caliber principle \cite{jaynes6,jaynes7}. Jaynes clearly stated that this does not represent a physical theory that explains the behavior of different systems by deductive reasoning from the first principles, but a form of statistical inference that makes predictions of observable phenomena from incomplete information \cite{jaynes6}. For this reason predictive statistical mechanics can not claim deductive certainty for its predictions. This does not mean that it ignores the laws of microphysics; it certainly uses everything known about the structure of microstates and any data on macroscopic quantities, without making any extra physical assumptions beyond what is given by available information. It is important to note that sharp, definite predictions of macroscopic behavior are possible only when it is characteristic of each of the overwhelming majority of microstates compatible with data. For the same reason, this is just the behavior that is reproduced experimentally under those constraints; this is known essentially as the principle of macroscopic uniformity \cite{jaynes1,jaynes2}, or reproducibility \cite{jaynes7}. In somewhat different context this property is recognized as the concept of macroscopic determinism, whose precise definition involves some sort of thermodynamic limit \cite{garrod}. 

In Jaynes' view, the dynamical invariance of the Gibbs' entropy gives a simple proof of the second law, which is then a special case of a general requirement for any macroscopic process to be experimentally reproducible \cite{jaynes4}. In the simple demonstration based on the Liouville theorem, this makes possible for Jaynes to generalize the second law beyond the restrictions of initial and final equilibrium states, by considering it a special case of a general restriction on the direction of any reproducible process \cite{jaynes4,jaynes8}. The real reason behind the second law, since phase space volume is conserved in the dynamical evolution, is a fundamental requirement on any reproducible process that the phase space volume $W^{\prime}$, compatible with the final state, can not be less than the phase space volume $W_0$ which describes our ability to reproduce the initial state \cite{jaynes4}. The arguments used in this demonstration imply also the question how to determine which nonequilibrium or equilibrium states can be reached from others, and this is not possible without information about dynamics, constants of motion, constraints, etc. The second law predicts only that a change of macroscopic state will go in the general direction of greater final entropy \cite{jaynes7}. Better predictions are possible only by introducing more information. Macrostates of higher entropy can be realized in overwhelmingly more ways, and this is the reason for high reliability of the Gibbs equilibrium predictions \cite{jaynes7}. In this context, Jaynes also speculated that accidental success in reversal of an irreversible process is exponentially improbable \cite{jaynes8}.   

Jaynes' interpretation of irreversibility and the second law reflects the point of view of the actual experimenter. Zurek \cite{zurek1} has introduced algorithmic randomness as the measure of the complexity of the microscopic state. He has prescribed entropy not only to the ensemble but also to the microscopic state. This prescription makes the principal distinction between his and Jaynes' approach. The basic laws of computation reflected in this interpretation allow Zurek to formulate thermodynamics from the point of view of Maxwell demon-type entities that can acquire information through measurements and process it in a manner analogous to Turing machines. According to Jaynes, the detailed description of microscopic development of the system can not be extracted from  the data about macroscopic development, and therefore it is not a subject of his approach. Increase of entropy is related to gradual decrease of information about possible microstates of the system compatible with data. The notion that the second law is a law of information dynamics, operating at the level of ``information bookkeeping", has been considered recently by Duncan and Semura \cite{dunsem1,dunsem2}. In this line of thinking, the dynamics of information is considered to be coupled, but fundamentally independent of energy dynamics.

MaxEnt algorithm and its methods represent a way of assigning probability distributions with the largest uncertainty and extent compatible with the available information, and for the same reasons, least biased with respect to unavailable information. Inferences drawn in this way depend only on our state of knowledge \cite{jaynes1,jaynes2}. In this work, two different applications of MaxEnt algorithm to macroscopic closed systems with Hamiltonian dynamics, and their time evolution, are examined in detail along with their consequences.  The concepts of {\it phase space paths} with the {\it path probability distribution} and associated {\it conditional probability distribution} are defined. The respective {\it path information entropy} and {\it conditional information entropy} are introduced in correspondence with definitions in Shannon's information theory \cite{shannon}. In the first approach, Liouville equation for the conditional probability distribution is introduced as a strict {\it microscopic constraint} on the time evolution in phase space, which is then completely determined by this constraint and initial values. Maximization of the conditional information entropy, subject to this constraint, predicts the macroscopic behavior that can happen in the greatest number of ways consistent with the information about microscopic dynamics. If probabilities are considered in the objective sense as a property of the system and not of our state of knowledge, full justification of this approach is possible only if our knowledge of the microscopic dynamics is complete. In a similar line of reasoning Grandy \cite{grandy2} has developed a detailed model of time dependent probabilities for macroscopic systems within MaxEnt formalism and applied it to typical processes in nonequilibrium thermodynamics and hydrodynamics \cite{grandy3,grandy4}. In a context of the interplay between macroscopic constraints on the system and its microscopic dynamics, it is interesting to note that MaxEnt has been also studied as a method of approximately solving partial differential equations governing the time evolution of probability distribution functions. For more complete further reference, we only mention here that this method, among other examples, has been applied to the Liouville--von Neumann equation \cite{tishby}, the family of dynamical systems with divergenceless phase space flows including Hamiltonian systems \cite{plastino1}, the generalized Liouville equation and continuity equations \cite{plastino2}. Universality of this approach has been established for the general class of evolution equations that conform to the essential requirements of linearity and preservation of normalization \cite{plastino3}. This method has been also considered for classical evolution equations with source terms within a framework where normalization is not preserved \cite{plastino4}.

The described first approach allows us to define concepts that are basis for our second approach. The difference is that Liouville equation for the conditional probability distribution is now introduced as a {\it macroscopic constraint}. This constraint on time evolution of the phase space probability density functions is now taken only on average, and it is given by the integral over accessible phase space. It is similar in this respect to constraints given by the data on macroscopic quantities. In Jaynes' predictive statistical mechanics more objectivity is ascribed to experimentally measured quantities than to probability distributions. The subjective aspect that becomes important here is that probabilities are assigned because of incomplete knowledge, i.e., partial information, and therefore represent our state of knowledge about the system. If information about dynamics is not sufficient to determine the time evolution, an average is taken over all cases possible on the basis of partial information. It is observed how elements of irreversible macroscopic behavior in closed systems with Hamiltonian dynamics are then a consequence of gradual loss of information about possible microstates of the system. This idea has been developed by Jaynes in the density-matrix formalism \cite{jaynes2}. In the approach which is developed here, we show that irreversible macroscopic behavior and Jaynes' interpretation based on reproducibility and information loss, has a clear mathematical description in the concepts of conditional information entropy, and its relation with the information entropy, i.e., in concepts of Shannon's information theory \cite{shannon}. At the end of the work, the subjective and objective aspects of both approaches are indicated, and relations with entropy production are discussed.

\section{Hamiltonian dynamics and phase space paths} \label{secHD} 
The dynamical state of a Hamiltonian system with $s$ degrees of freedom is described by the coordinates $q_1, q_2, \dots , q_s$ and the momenta $p_1, p_2, \dots , p_s$. At any time $t$ it is represented by a point in the $2s$-dimensional Euclidean space $\Gamma $ here for our purposes called the phase space of the system. The notation $(q,p)$ is introduced for the set of $2s$ coordinates and momenta. The time dependence of $2s$ dynamical variables $(q,p)$ is determined by Hamilton's equations  
\begin{equation}
\dot q_i = \frac{\partial H}{\partial p_i} , \qquad \qquad \dot p_i =  - \frac{\partial H}{\partial q_i} , \qquad \qquad1 \leq i \leq s , \label{eq1}
\end{equation} 
where $H \equiv H(q,p)$ is the Hamiltonian function of the system. Given the values $(q_0,p_0)$ at some time $t_0$, the solution of Hamilton's equations (\ref{eq1}) uniquely determines the values of dynamical variables $(q,p)$ at any other time $t$, 
\begin{equation}
q_i = q_i(t; q_0,p_0) , \qquad \qquad p_i = p_i(t; q_0,p_0) , \qquad \qquad1 \leq i \leq s . \label{eq2} 
\end{equation}   
Any point $(q,p)$ in the phase space $\Gamma $ describes a curve called a {\it phase space path}, given by the uniquely determined solution of (\ref{eq1}). At any time $t$ through each point of $\Gamma $ passes only one path, and this is denoted by the index in $(q,p)_\omega $, where $\omega \in \Omega (\Gamma )$. The set $\Omega (\Gamma )$ is the set of all paths in $\Gamma $. The velocity $v$ of the point $(q,p)$ in the phase space $\Gamma $ at time $t$ is given by 
\begin{equation}
v \equiv  \vert {\bf v}\vert  = \sqrt {\sum _{i=1}^{s} \left (\frac {dq_{i}}{dt}\right ) ^2 + \sum _{i=1}^{s}\left (\frac{dp_{i}}{dt}\right ) ^2} = \sqrt {\sum _{i=1}^{s} \left (\frac {\partial H}{\partial p_i}\right ) ^2 + \sum _{i=1}^{s}\left (\frac{\partial H}{\partial q_i}\right ) ^2}. \label{eq2a}
\end{equation}
The velocity vector ${\bf v}((q,p)_\omega , t)$ is tangential at the point  $(q,p)_\omega  \in \Gamma $ to the phase space path $\omega $ passing through it at time $t$. For the systems considered here the Hamiltonian function $H(q,p)$ does not depend on time and the velocity field ${\bf v}(q,p, t)$ of all points in the phase space $\Gamma $ is stationary, i.e., ${\bf v}(q,p, t) = {\bf v}(q,p)$.

Let $M_0$ be any measurable (in the sense of Lebesgue) set of points in the phase space $\Gamma $. In the Hamiltonian motion the set $M_0$ is transformed into another set $M_t$ during an interval of time $t$. Liouville's theorem asserts that the measure of the set $M_t$ for any $t$  coincides with the measure of the set $M_0$ \cite[pp. 15--16]{khinchin}. This theorem proves that the measure in the phase space $\Gamma $,
\begin{equation}
\mu (M_t) = \int _{M_t}dq_1\dots dq_s dp_1\dots dp_s = \int _{M_t}d\Gamma ,\label{eq3}  
\end{equation}
is invariant under Hamiltonian motion. In the notation used in (\ref{eq3}), volume element $dq_1\dots dq_s dp_1\dots dp_s$ of the phase space $\Gamma $ is denoted by $d\Gamma $. One immediate corollary \nopagebreak \cite[pp. 18--19]{khinchin} of Liouville's theorem is that, if $M_0$ is a Lebesgue measurable set of points of the phase space $\Gamma $, of finite measure, and $f(q,p)$ a phase function Lebesgue integrable over $\Gamma $, then 
\begin{equation}
\int _{M_t} f(q,p)d\Gamma = \int _{M_0} f(q(t; q_0,p_0),p(t; q_0,p_0))d\Gamma_0 . \label{eq4} 
\end{equation}
Equation (\ref {eq4}) is obtained by changing the variables in the integral and introducing new variables $(q_0,p_0)$, related to the variables $(q,p)$ by transformation of the space $\Gamma $ into itself in Hamiltonian motion, given by (\ref{eq2}). If, in particular, the set $M$ is invariant to the Hamiltonian motion, then using this corollary, it is easy to show how an integral of a phase function $f(q,p)$ over the invariant set $M$ is transformed into an integration over the set $\Omega (M)$ of all paths in $M$. This procedure is now developed in the rest of this section. It is used in the definition of probability distributions in Sect. \ref{sec_Mic}.  

At any time $t$ through each point $(q,p)_\omega  \in \Gamma $ passes only one path $\omega \in \Omega (\Gamma )$, that also passes through the point $(q_0,p_0)_\omega \in \Gamma $ given by the inverse of (\ref{eq2}). The infinitesimal volume element $d\Gamma _0$ around the point $(q_0,p_0)_\omega$ can be written as $d\Gamma _0 = ds_{0 \omega } dS_{0 \omega }$. Here, $ds_{0 \omega}$ is the infinitesimal distance along the path $\omega $. The infinitesimal element $dS_{0 \omega }$ of the surface $S_0(M)$ intersects the path $\omega $ perpendicularly at the point $(q_0,p_0)_\omega$. The surface $S_0(M)$ is perpendicular to all paths in the set $\Omega (M)$ of paths in $M$. 

The invariance of the measure $d\Gamma $ to Hamiltonian motion and the fact that the velocity field ${\bf v}(q,p)$ in $\Gamma $ is stationary as the Hamiltonian function $H(q,p)$ does not depend on time, lead to the following consequence. For any phase space path $\omega \in \Omega (\Gamma )$, the product of the velocity $v((q,p)_\omega )$ and the infinitesimal surface $dS_{\omega }$ intersecting the path $\omega $ perpendicularly at the point $(q,p)_\omega $, is constant under Hamiltonian motion along the entire length of the path $\omega $, i.e., 
\begin{equation}
v ((q,p)_\omega ) dS_\omega  = const . \label{eq4b}
\end{equation}
For any two points $(q_0,p_0)_\omega $ and $(q_a,p_a)_\omega $ on the same path $\omega $, the following relation is obtained from (\ref{eq4b}):
\begin{equation}
v ((q_0,p_0)_\omega ) dS_{0 \omega}  = v ((q_a,p_a)_\omega ) dS_{a \omega} . \label{eq4c}
\end{equation}
The infinitesimal element $dS_{a \omega }$ of the surface $S_a(M)$ intersects the path $\omega $ perpendicularly at the point $(q_a,p_a)_\omega$. Like the surface $S_0(M)$, surface $S_a(M)$ is also perpendicular to all paths in $\Omega (M)$. The infinitesimal elements $dS_{0 \omega }$ and $dS_{a \omega}$ of the two surfaces $S_0(M)$ and $S_a(M)$ are connected by the path $\omega $ and neighboring paths determined by solutions of Hamilton's equations. The integral over surface $S_a(M)$ is transformed using (\ref{eq4c}) into integration over surface $S_0(M)$,
\begin{equation}
\int _{S_a(M)} dS _{a \omega } = \int _{S_0(M)} \frac{v((q_0,p_0)_\omega )}{v ((q_a,p_a)_\omega)} dS _{0 \omega } .  \label{eq4d}
\end{equation}
Functional dependence between the points $(q_0,p_0)_\omega $ and $(q_a,p_a)_\omega $ on the path $\omega $ is  not explicitly written in the integral (\ref{eq4d}); it is implied that this functional dependence is determined from solutions of Hamilton's equations.

The following notation is introduced by using (\ref{eq2}), with the times $t$ and $t_0$ fixed, in the phase function $f(q,p)$:
\begin{equation}
f(q(t; q_0,p_0),p(t; q_0,p_0)) \equiv g(q_0,p_0, t_0) . \label{eq5a}
\end{equation}
Equation (\ref{eq5a}) is then substituted (with $t$ and $t_0$ fixed and the indices in $(q_0,p_0)$ replaced by the indices $(q_a,p_a)$) into the integral (\ref{eq4}), taken over the set $M$ which is invariant to Hamiltonian motion. This leads to following equality:
\begin{equation}
\int _M f(q,p)d\Gamma = \int _{M}  g(q_a,p_a, t_0)d\Gamma _a  . \label{eq5}
\end{equation}
The integral (\ref{eq5}) is then transformed using relation (\ref{eq4c}) and $d\Gamma _a = ds_{a \omega } dS_{a \omega }$: 
\begin{equation}
\int _{M} g(q_a,p_a, t_0)ds_{a \omega } dS _{a \omega } = \int _{S_0(M)} dS _{0 \omega }v((q_0,p_0)_\omega )\int _\omega \frac{g(q_a,p_a, t_0)}{v (q_a,p_a)} ds_{a \omega } . \label{eq6a}
\end{equation}
The function
\begin{equation}
F((q_0,p_0)_\omega ,t_0) = v((q_0,p_0)_\omega )\int _\omega \frac{g(q_a,p_a, t_0)}{v(q_a,p_a)} ds_{a \omega }, \label{eq6}
\end{equation}
is defined on the surface $S_0(M)$ and is called a {\it path function} or {\it path distribution}. The integral in the relation (\ref{eq6}) defining a path function $F((q_0,p_0)_\omega ,t_0)$ is over the entire length of the path $\omega $ intersected perpendicularly by the surface $S_0(M)$ at the point $(q_0,p_0)_\omega$. Infinitesimal element of the phase space path $\omega $ passing through the point $(q_a,p_a)_\omega$ is $ds_{a \omega }$, and the time $t_0$ in the integral (\ref{eq6}) is fixed.   

If the phase function $f(q,p)$ in (\ref{eq5}) is a phase space probability density function, equal to zero everywhere outside the invariant set $M$, it is straightforward to prove that the path function $F((q_0,p_0)_\omega ,t_0)$ defined by (\ref{eq6}) satisfies the nonnegativity and normalization conditions required from probability distributions. Nonnegativity and normalization of the function $F((q_0,p_0)_\omega ,t_0)$ which then represents a {\it path probability distribution}, follow from the nonnegativity and normalization properties of the related phase space probability density function $f(q,p)$. With the help of (\ref{eq5}) and (\ref{eq6a}) and the definition of $F((q_0,p_0)_\omega ,t_0)$ in (\ref{eq6}), one then obtains the normalization property
\begin{equation}
\int _M f(q,p)d\Gamma = \int _{S_0(M)} F((q_0,p_0)_\omega ,t_0) dS_0 = 1 . \label{eq8} 
\end{equation}
Nonnegativity is established for all $(q_0,p_0)_\omega  \in S_0(M) $ in a similar way. Integral over any invariant and measurable subset of the set $M $ is transformed, in the way described above, into integral over a corresponding measurable subset on the surface $S_0(M) $. It is clear also that the measure defined on the surface $S_0(M) $ can be utilized as a measure on the set $\Omega (M)$ of all phase space paths in some invariant set $M $. The correspondence between points  $(q_0,p_0)_\omega  \in S_0(M) $ and paths $\omega \in \Omega (M)$ is one-to-one.

\section{Microstate probability and path probability} \label{sec_Mic}
It is now possible to relate the microstate probability and the path probability in the phase space $\Gamma $ of the system. Let the function $f(q,p,t)$ be a microstate probability density function on $\Gamma $. All points in the phase space $\Gamma $ move according to Hamilton's equations (\ref{eq1}) and $f(q,p,t)$ satisfies the Liouville equation
\begin{equation}
{\partial f \over \partial t} + \sum _{i=1} ^s \left ({\partial f \over \partial q_i}{\partial H \over \partial p_i} - {\partial f \over \partial p_i}{\partial H \over \partial q_i}\right ) \equiv {df \over dt} = 0 .\label{eq9}
\end{equation}   
Since $df/dt$ is a total or hydrodynamic derivative, (\ref{eq9}) expresses that the time rate of change of $f(q,p,t)$ is zero along any phase space path given by uniquely determined solution of Hamilton's equations. In the notation used here, this fact is written as
\begin{equation}
f((q,p)_\omega ,t) = f((q_0,p_0)_\omega ,t_0) , \label{eq10}
\end{equation}
where points on the path $\omega \in \Omega (\Gamma )$ are related by (\ref{eq2}).

In addition to the definition of the path probability distribution $F((q_0,p_0)_\omega ,t_0)$ via microstate probability density function $f(q,p,t)$, it is possible to give another equivalent definition of $F((q_0,p_0)_\omega ,t_0)$. In order to accomplish this, probability density function $\mathcal {F}(q,p,t;q_0,p_0 ,t_0)$ is introduced on the $4s$-dimensional Euclidean space $\Gamma \times \Gamma $. This function has the following special properties. If the integral of $\mathcal {F}(q,p,t;q_0,p_0 ,t_0)$ is taken over the phase space $\Gamma $ with $(q,p)$ as the integration variables, it gives the microstate probability density function $f(q_0,p_0,t_0)$ at time $t_0$, 
\begin{equation}
\int _\Gamma \mathcal{F}(q,p,t;q_0,p_0 ,t_0) d\Gamma = f(q_0,p_0, t_0) .  \label{eq11}
\end{equation}
Microstate probability density function $f(q,p,t)$ at time $t$ is obtained analogously,
\begin{equation}
\int _\Gamma \mathcal{F}(q,p,t;q_0,p_0 ,t_0) d\Gamma_0 = f(q,p, t) . \label{eq12}
\end{equation}
It is straightforward to prove, using relation (\ref{eq10}), that (\ref{eq11}) and (\ref{eq12}) are satisfied if the function $\mathcal{F}(q,p,t;q_0,p_0 ,t_0)$ has the following form:
\begin{equation}
\mathcal{F}(q,p,t; q_0,p_0 ,t_0) = f(q,p, t)\prod_{i=1}^{s}\delta (q_i - q_i(t;q_0,p_0))\delta (p_i - p_i(t;q_0,p_0)) , \label{eq13}
\end{equation}
where $q_i(t;q_0,p_0)$ and $p_i(t;q_0,p_0)$ are given by (\ref{eq2}) and $\delta$-s are Dirac delta functions. In the space $\Gamma \times \Gamma $ function $\mathcal{F}(q,p,t;q_0,p_0 ,t_0)$ given by (\ref{eq13}) represents the probability density that the point corresponding to the state of the system is in the element $d\Gamma _0$ around the point $(q_0,p_0)$ at time $t_0$ and in the element $d\Gamma $ around the point $(q,p)$ at time $t$.

As explained in (\ref{eq12}), microstate probability density function $f(q,p, t)$ is given by the integral of the function $\mathcal{F}(q,p,t;q_0,p_0 ,t_0)$ over $\Gamma $ with $(q_0,p_0)$ as integration variables. Now, we assume that the set $M$ of all points in $\Gamma $ which represent possible microstates of the system is invariant to Hamiltonian motion. By applying the similar procedure and notation that was already introduced in relations (\ref{eq5}) and (\ref{eq6a}), the integral (\ref{eq12}) can now be written as  
\begin{equation}
f(q,p, t) = \int _{S_0(M)}dS_{0 \omega }v((q_0,p_0)_\omega )\int _\omega  \frac{\mathcal{F}(q,p,t;q_a,p_a ,t_0)}{v (q_a,p_a)} ds_{a \omega } . \label{eq15}
\end{equation}
Along with the lines leading to (\ref{eq15}), the function $G(q,p,t; (q_0,p_0)_\omega ,t_0 )$ is also introduced:
\begin{equation}
G(q,p,t; (q_0,p_0)_\omega ,t_0 ) = v((q_0,p_0)_\omega )\int _\omega  \frac{\mathcal{F}(q,p,t; q_a,p_a ,t_0)}{v (q_a,p_a)} ds_{a \omega } . \label{eq16}
\end{equation}
The integral in the definition of $G(q,p,t; (q_0,p_0)_\omega ,t_0 )$ in (\ref{eq16}) is over the entire length of the phase space path $\omega $ intersected perpendicularly by the surface $S_0(M)$ at the point $(q_0,p_0)_\omega$. Using (\ref{eq16}), relation (\ref{eq15}) is then written as  
\begin{equation}
f(q,p, t) = \int _{S_0(M)}G(q,p,t; (q_0,p_0)_\omega ,t_0 )dS_0 .  \label{eq17}
\end{equation}
It is clear that the expression 
\begin{equation}
G(q,p,t; (q_0,p_0)_\omega ,t_0 )dS_0d\Gamma \equiv \mathrm{P}(q,p , t  \cap  (q_0,p_0)_\omega , t_0) , \label{eq18}
\end{equation}
represents the probability that the point corresponding to the state of the system is at time $t_0$ anywhere along the paths which pass through an infinitesimal element $dS_0$ around $(q_0,p_0)$ on the surface $S_0(M)$, and that at some different time $t$ it is in the volume element $d\Gamma $ around $(q,p)$. 

Another definition of the path probability distribution $F((q_0,p_0)_\omega ,t_0)$, in addition to (\ref{eq6}), is now possible in this way. It is given by the integral  
\begin{equation}
F((q_0,p_0)_\omega ,t_0) = \int_\Gamma  G(q,p,t; (q_0,p_0)_\omega ,t_0 )d\Gamma . \label{eq19}
\end{equation}
Then, in accordance with the theory of probability, the ratio 
\begin{equation}
{G(q,p,t;(q_0,p_0)_\omega ,t_0) dS_0 d\Gamma \over F((q_0,p_0)_\omega ,t_0) dS _0} \equiv \mathrm{P} (q,p , t \vert (q_0,p_0)_\omega , t_0), \label{eq21} 
\end{equation}
represents the {\it conditional probability} that at time $t$ the point corresponding to the state of the system is in the element $d\Gamma $ around $(q,p)$, if at time $t_0$ it is anywhere along the paths passing through the infinitesimal element $dS_0$ around $(q_0,p_0)$ on the surface $S_0(M)$. Relation (\ref{eq19}) then proves that the integral of (\ref{eq21}) over $\Gamma $ satisfies the normalization condition, i.e.,
\begin{equation}
\int _\Gamma { G(q,p,t;(q_0,p_0)_\omega ,t_0) dS _0 \over F((q_0,p_0)_\omega ,t_0) dS _0}d\Gamma   = 1 . \label{eq22} 
\end{equation}
To set up all the tools of probability theory needed in this work, conditional probability distribution $D(q,p , t \vert (q_0,p_0)_\omega , t_0)$ that corresponds to conditional probability (\ref{eq21}), is defined by the relation
\begin{equation}
D(q,p , t \vert (q_0,p_0)_\omega , t_0) = {G(q,p,t;(q_0,p_0)_\omega ,t_0) \over F((q_0,p_0)_\omega ,t_0) } . \label{eq23} 
\end{equation}

The relation (\ref{eq21}), like the relation (\ref{eq18}), represents probability which is conserved in the phase space $\Gamma $. The total time derivative (i.e., time rate of change along the Hamiltonian flow lines) of this probability is equal to zero. In the relation (\ref{eq21}) for the conditional probability, the path probability distribution $F((q_0,p_0)_\omega ,t_0)$ and the surface element $dS _0$ are independent of the variables $t$ and $(q,p)$. Also, measure $d\Gamma $ is invariant to Hamiltonian motion. Therefore, it follows that the total time derivative of the conditional probability (\ref{eq21}) is equal to zero if and only if 
\begin{equation}
{dG \over dt} \equiv {\partial G \over \partial t} + \sum _{i=1} ^s \left ({\partial G \over \partial q_i}{\partial H \over \partial p_i} - {\partial G \over \partial p_i}{\partial H \over \partial q_i}\right )  = 0\ . \label{eq24}
\end{equation} 
This is a straightforward demonstration that the probability distribution $G(q,p,t;(q_0,p_0)_\omega ,t_0)$ satisfies the equation analogous to the Liouville equation (\ref{eq9}) for the microstate probability distribution $f(q,p,t)$.

\section{Information entropies and MaxEnt algorithm} \label{secIE}

The quantity of the form  $H = -\sum_i p_i\log p_i$ has a central role in information theory as a measure of information, choice and uncertainty for different probability distributions $p_i$. In an analogous manner Shannon \cite{shannon} has defined entropy of a continuous distribution and entropy of $N$-dimensional continuous distribution. As pointed out by Jaynes \cite{jaynes3}, the analog of $-\sum_{i=1}^{n} p_i\log p_i$ for a discrete probability distribution $p_i$ which goes over in a limit of infinite number of points into a continuous distribution $w(x)$ (in such a way that the density of points divided by the total number of points approaches a definite function $m(x)$) is given by 
\begin{equation}
S_I = -\int w(x)\log\left [\frac{w(x)}{m(x)}\right ]dx . \label{eq25}
\end{equation}
Shannon assumed the analog $-\int w(x)\log[w(x)]dx$, but he also pointed out an important difference between his definitions of discrete and continuous entropies. If we change coordinates, the entropy of a continuous distribution will in general change in the way taken into account by Shannon \cite{shannon}. To achieve the required invariance of entropy of a continuous distribution under a change of the independent variable, it is necessary to introduce the described modification that follows from mathematical deduction \cite{jaynes3}. This is achieved with an introduction of the measure function $m(x)$ and yields the invariant information measure (\ref{eq25}). If a uniform measure $m=const$ is assumed \cite{jaynes3}, the invariant information measure (\ref{eq25}) differs from the Shannon's definition of entropy of a continuous distribution \cite{shannon} by an irrelevant additive constant.   

Shannon \cite{shannon} has also defined the joint and conditional entropies of a joint distribution of two continuous variables (which may themselves be multidimensional), concepts that are applied in this work. In the previous section, {\it joint distribution} $G(q,p,t;(q_0,p_0)_\omega ,t_0)$ of two continuous multidimensional variables $(q,p) \in \Gamma $ and $(q_0,p_0)_\omega  \in S_0(M)$ was introduced. Following the detailed explanation of (\ref{eq18}), $G(q,p,t;(q_0,p_0)_\omega ,t_0)dS_0d\Gamma $ represents the probability of the joint occurrence of two events: the first occurring at time $t_0$ among the set of all possible phase space paths $\Omega (M)$ and the second occurring at time $t$ among the set of all possible phase space points $M$ which is invariant to Hamiltonian motion. In accordance with Shannon's definition \cite{shannon}, {\it joint information entropy} of the joint distribution $G(q,p,t;(q_0,p_0)_\omega ,t_0)$ is given by
\begin{equation}
S_{I}^{G}(t, t_0) = - \int _{S_0(M)} \int_\Gamma G \log G \ d\Gamma dS _0  . \label{eq26}
\end{equation} 
The notation $S_{I}^{G}(t, t_0)$ indicates that it is a function of times $t$ and $t_0$, through the distribution $G \equiv  G(q,p,t;(q_0,p_0)_\omega ,t_0)$. Following Shannon's definition \cite{shannon}, {\it conditional information entropy} of the joint distribution $G(q,p,t;(q_0,p_0)_\omega ,t_0)$ is then given by
\begin{equation}
S_{I}^{DF} (t, t_0) = - \int _{S_0(M)} \int_\Gamma G\log \left[\frac{G}{F}\right ] \ d\Gamma dS _0  , \label{eq27}
\end{equation}
where $F \equiv  F((q_0,p_0)_\omega ,t_0 )$ is the path probability distribution. Using the definition of $D(q,p,t | (q_0,p_0)_\omega ,t_0 )$ in (\ref{eq23}), one immediately obtains the equivalent form of the conditional information entropy (\ref{eq27}): 
\begin{equation}
S_{I}^{DF} (t, t_0) = - \int _{S_0(M)} \int_\Gamma DF\log D \ d\Gamma dS _0 . \label{eq28}
\end{equation}
From (\ref{eq28}) it is clear that the conditional information entropy $S_{I}^{DF} (t, t_0)$ is the average of the entropy of conditional probability $D(q,p,t | (q_0,p_0)_\omega ,t_0 )$, weighted over all possible phase space paths $\omega \in \Omega (M)$ according to the path probability distribution $F((q_0,p_0)_\omega ,t_0 )$. 

Relation between the information entropies $S_{I}^{G}(t, t_0)$ and $S_{I}^{DF} (t, t_0)$, introduced in (\ref{eq26}) and (\ref{eq27}), is completed by introducing the information entropy of the distribution $F((q_0,p_0)_\omega ,t_0 )$, or alternatively, {\it path information entropy}: 
\begin{equation}
S_{I}^{F} (t_0) = - \int _{S_0(M)} F\log F \ dS _0 . \label{eq29} 
\end{equation}
Relation between $S_{I}^{G}(t, t_0)$, $S_{I}^{DF} (t, t_0)$ and $S_{I}^{F} (t_0)$ is obtained straightforwardly, using (\ref{eq23}) in (\ref{eq26}), and then applying the properties of probability distributions. In this way one obtains
\begin{equation}
S_{I}^{G}(t,t_0) = S_{I}^{DF}(t,t_0) + S_{I}^{F} (t_0) . \label{eq30}
\end{equation}
Relation (\ref{eq30}), in accordance with the analogous relation of Shannon \cite{shannon}, asserts that the uncertainty (or entropy) of the joint event is equal to the uncertainty of the first plus the uncertainty of the second event when the first is known. 

It is important to give some additional comments to (\ref{eq30}). In general, uncertainty of the joint event is less then or equal to the sum of uncertainties of the two individual events, with the equality if (and only if) the two events are independent \cite{shannon}. The probability distribution of the joint event is given here by $G(q,p,t;(q_0,p_0)_\omega ,t_0)$. Information entropy or uncertainty of one of them (in this case called the second event because of its occurrence at a later time) is equal 
\begin{equation}
S_{I}^{f} (t) = - \int_\Gamma f\log f \ d\Gamma . \label{eq31}
\end{equation}
The quantity $S_{I}^{f} (t)$ is the information entropy of the microstate probability distribution $f(q,p,t)$, or in short, {\it information entropy}. The uncertainty of the first event is given by the path information entropy $S_{I}^{F} (t_0)$ defined in (\ref{eq29}). The aforementioned property of information entropies is given here for $S_{I}^{G}(t,t_0)$, $S_{I}^{f} (t)$ and $S_{I}^{F} (t_0)$ by the following relation: 
\begin{equation}
S_{I}^{G}(t,t_0) \leq  S_{I}^{f} (t) + S_{I}^{F} (t_0) , \label{eq32}
\end{equation}
with the equality if (and only if) the two events are independent. Furthermore, from (\ref{eq30}) and (\ref{eq32}), one obtains an important relation between $S_{I}^{f} (t)$ and $S_{I}^{DF}(t,t_0)$:
\begin{equation}
S_{I}^{f} (t) \geq  S_{I}^{DF}(t,t_0) , \label{eq33}
\end{equation}
with the equality if (and only if) the two events are independent.

In terms of probability, the events occurring at time $t_0$ among the set of all possible phase space paths $\Omega (M)$ and at any time $t$ among the set of all possible phase space points $M \subset \Gamma $, are not independent. If we assume that the values of joint probability distribution $G(q,p,t;(q_0,p_0)_\omega ,t_0) $ are physically well defined (in the sense of (\ref{eq18})) for all points $(q,p) \in \Gamma $ and $(q_0,p_0) \in S_0(M)$ at given initial time $t = t_0$, its values are then determined at all times $t$ in the entire phase space $\Gamma$ via the Liouville equation (\ref{eq24}). Simple deduction leads to the conclusion that maximization of the conditional information entropy $S_{I}^{DF} (t, t_0)$, subject to the constraints of Liouville equation (\ref{eq24}) and normalization, can not yield the upper bound which is given (at any time $t$) by the value of the information entropy $S_{I}^{f} (t)$ in (\ref{eq33}). Attaining this upper bound would require statistical independence or, in other words, a complete loss of correlation between the set of possible phase space paths $\Omega (M)$ at time $t_0$ and the set of possible phase space points $M \subset \Gamma $ at time $t$. This is precluded at any time $t$ by the constraint implied by the Liouville equation (\ref{eq24}), and the requirement that the joint probability distribution $G(q,p,t;(q_0,p_0)_\omega ,t_0) $ is well defined.

At this point it is helpful to make a distinction between two aspects of time evolution. The first is a microscopic aspect which represents a problem of dynamics implied in this work by Hamilton's equations. The solutions are represented in $\Gamma $ as phase space paths. Predicting macroscopic time evolution represents a problem of available information and inferences from that partial information. Therefore, microscopic dynamics and the respective phase space paths are also part of this problem of incomplete information. In a case of macroscopic system, information about microscopic dynamics is very likely to be incomplete for variety of different possible reasons. However, in the absence of more complete knowledge, Hamilton's equations (\ref{eq1}) and the set of possible phase space paths are the representation of our information about microscopic dynamics. It is natural to assume that the predicted macroscopic time evolution for a closed system is consistent with our knowledge of microscopic dynamics, even when this knowledge is not complete. 

All arguments mentioned before lead to the conclusion that regarding Liouville equation (\ref{eq24}) as a strict {\it microscopic constraint} on time evolution is equivalent to having complete information about microscopic dynamics. Following previously introduced assumptions, the Liouville equation (\ref{eq24}) can also be regarded as a {\it macroscopic constraint} on time evolution. If our information about microscopic dynamics is not sufficient to determine the time evolution, an average is taken over all cases possible on the basis of our partial information. The conditional information entropy $S_{I}^{DF} (t, t_0)$ is then maximized subject to the constraint of Liouville equation (\ref{eq24}), introduced as a phase space average, or more precisely, an integral over phase space similarly to other macroscopic constraints. In predictive statistical mechanics formulated by Jaynes, inferences are drawn from probability distributions whose sample spaces represent what is known about the structure of microstates, and maximize information entropy subject to the available macroscopic data \cite{jaynes6}. In this way ``objectivity" of probability assignments and predictions is ensured from introducing extraneous assumptions not warranted by data. In this work we introduce the same basic idea into stage (2) of the problem of prediction for closed Hamiltonian systems. This approach allows us to consider the incomplete nature of our information about microscopic dynamics in a rational way, and leads to the loss of correlation between the initial phase space paths and final microstates and to uncertainty in prediction. The conditional information entropy $S_{I}^{DF} (t, t_0)$ is the measure of this uncertainty, related to loss of information about the state of the system.

\section{MaxEnt inferences and time evolution}

In the first approach, time evolution of the conditional probability distribution $D(q,p,t | (q_0,p_0)_\omega ,t_0 )$ in the interval $t_0 \leq t \leq t_a$ should satisfy the following constraints: normalization condition  
\begin{equation}
\int _M D(q,p,t | (q_0,p_0)_\omega ,t_0 )d\Gamma  = 1 , \label{eq37}
\end{equation}
and the Liouville equation for $D(q,p,t | (q_0,p_0)_\omega ,t_0 )$,
\begin{equation}
{\partial D \over \partial t} + \sum _{i=1} ^s \left ({\partial D \over \partial q_i}{\partial H \over \partial p_i} - {\partial D \over \partial p_i}{\partial H \over \partial q_i}\right ) = 0 . \label{eq36}
\end{equation} 
From (\ref{eq23}) it follows that the constraints given by (\ref{eq24}) and (\ref{eq36}) are equivalent. By definition, the set of all possible microstates $M \subset \Gamma$ is an invariant set. The normalization constraint (\ref{eq37}) contains information about the structure of possible microstates in $\Gamma $, in the time interval under consideration $t_0 \leq t \leq t_a$. Information about microscopic dynamics is represented by the set $\Omega (M)$ of possible phase space paths in $\Gamma$. In addition, this information is contained in the Liouville equation (\ref{eq36}). The assigned path probability distribution $F((q_0,p_0)_\omega ,t_0 )$ is compatible with the available information. 

Time derivative of the conditional information entropy $S_{I}^{DF}(t,t_0)$ in (\ref{eq28}) is given by 
\begin{equation}
\frac {dS_{I}^{DF}(t,t_0)}{dt} = - \int _{S_0(M)} \int_M \frac{\partial D}{\partial t} F\log D \ d\Gamma dS _0 -  \int _{S_0(M)} \int_M \frac{\partial D}{\partial t} F \ d\Gamma dS _0 . \label{eq38}
\end{equation}
Because of the normalization, (\ref{eq37}), the last term in (\ref{eq38}) is equal to zero. At time $t_a$, conditional information entropy $S_{I}^{DF}(t_a,t_0)$ is given by the expression, 
\begin{equation}
S_{I}^{DF} (t_a, t_0) = - \int _{t_0}^{t_a} \int _{S_0(M)} \int_M \frac{\partial D}{\partial t}F\log D \ d\Gamma dS _0 dt + S_{I}^{DF}(t_0, t_0) . \label{eq40}
\end{equation}
The following functional is then formed 
\begin{equation}
J[D] = S_{I}^{DF} (t_a, t_0) - S_{I}^{DF}(t_0, t_0) =  \int _{t_0}^{t_a} \int _{S_0(M)} \int_M  K(D, \partial _t D) d\Gamma dS _0 dt , \label{eq41}
\end{equation}
with the function $K(D, \partial _t D)$ given by
\begin{equation}
K(D, \partial _t D) = - \frac{\partial D}{\partial t}F \log D . \label{eq42}
\end{equation} 
In the variational problem which is considered here, functional $J[D]$ in (\ref{eq41}) is rendered stationary with respect to variations subject to the constraints (\ref{eq37}) and (\ref{eq36}). On the boundary of integration region $M \times (t_0,t_a)$ in the integral (\ref{eq41}), function $D(q,p,t | (q_0,p_0)_\omega ,t_0 )$ is not required to take on prescribed values. The constraints given by (\ref{eq37}) and (\ref{eq36}) are written here in equivalent but more suitable form: 
\begin{eqnarray}
\varphi_1 ((q_0,p_0)_\omega ,t_0; t, D) = F \int_M D \ d\Gamma   -  F = 0 , \label{eq43}
\end{eqnarray}
and
\begin{eqnarray}
\varphi _2 ((q_0,p_0)_\omega ,t_0; q,p ,t, \partial _q D, \partial _p D, \partial _t D) = \left [{\partial D \over \partial t} + \sum _{i=1} ^s \left ({\partial D \over \partial q_i}{\partial H \over \partial p_i} - {\partial D \over \partial p_i}{\partial H \over \partial q_i}\right )\right ]F = 0 . \label{eq44}
\end{eqnarray} 

Methods for variational problems with this type of constraints exist and one can develop them and apply in practical problems \cite{wan}. Here, in the notation which is adapted to this particular problem, the following functionals are introduced:      
\begin{equation}
C_1[D, \lambda _1] = \int_{S_0(M)} \int _{t_0}^{t_a} \lambda _1 \varphi_1 \ dt dS_0  , \label{eq45}
\end{equation} 
and
\begin{equation}
C_2[D, \lambda _2] = \int _{S_0(M)} \int _{t_0}^{t_a} \int_M \lambda _2 \varphi _2 \ d\Gamma dt dS_0  . \label{eq46}
\end{equation} 
The Lagrange multipliers $\lambda _1 \equiv \lambda _1 ((q_0,p_0)_\omega ,t_0; t)$ and $\lambda _2 \equiv \lambda _2 ((q_0,p_0)_\omega ,t_0; q,p ,t)$  are functions defined in the integration regions in (\ref{eq45}) and (\ref{eq46}). For any function with continuous first partial derivatives, Euler equation for the constraint $\varphi _2 \equiv \varphi _2 ((q_0,p_0)_\omega ,t_0; q,p ,t, \partial _q D, \partial _p D, \partial _t D)$ is equal to zero. Following the multiplier rule  for such problems explained in ref. \cite{wan}, we introduce an additional (constant) Lagrange multiplier $\lambda _0$ as a multiplicative factor for $K$, 
\begin{equation} 
J[D, \lambda_0] = \int _{t_0}^{t_a} \int _{S_0(M)} \int_M  \lambda _0 K(D, \partial _t D) \ d\Gamma dS _0 dt . \label{eq47}
\end{equation}
The functional $I[D, \lambda_0, \lambda_1, \lambda_2 ]$ is formed from $J[D, \lambda_0]$, $C_1[D, \lambda _1]$ and $C_2[D, \lambda _2]$:  
\begin{equation}
I[D, \lambda_0, \lambda_1, \lambda_2 ] = J[D, \lambda_0] - C_1[D, \lambda _1] - C_2[D, \lambda _2] \label{eq48} . 
\end{equation} 
Existence of Lagrange multipliers $\lambda_0 \ne 0$, $\lambda_1$ and $\lambda_2 $, such that the variation of $I[D, \lambda_0, \lambda_1, \lambda_2 ]$ is stationary  $\delta I = 0$, represents a proof that it is possible to make $J[D]$ in (\ref{eq41}) stationary subject to constraints (\ref{eq43}) and (\ref{eq44}). The function $D(q,p,t | (q_0,p_0)_\omega ,t_0 )$ which renders $J[D]$ stationary subject to (\ref{eq43}) and (\ref{eq44}) must satisfy the Euler equation:   
\begin{eqnarray}
&& \lambda _0 \left \{\frac{\partial K}{\partial D} - \frac{d}{dt}\left (\frac{\partial K}{\partial (\partial _t D)}\right ) - \sum_ {i = 1}^s \left [\frac{d}{dq_i}\left (\frac{\partial K}{\partial (\partial _{q_i} D)}\right ) + \frac{d}{dp_i}\left (\frac{\partial K}{\partial (\partial _{p_i} D)}\right ) \right ]\right \} \cr\nonumber\\
&& - \lambda _1F + \left [{\partial \lambda _2 \over \partial t} + \sum _{i=1} ^s \left ({\partial \lambda _2 \over \partial q_i}{\partial H \over \partial p_i} - {\partial \lambda _2 \over \partial p_i}{\partial H \over \partial q_i}\right )\right ]F   = 0 . \label{eq50}
\end{eqnarray} 
It is easy to check that the term multiplied by $\lambda _0$ in Euler equation (\ref{eq50}) is equal to zero. Stationarity of the functional $I[D, \lambda_0, \lambda_1, \lambda_2 ]$ in (\ref{eq48}) is therefore possible even with $\lambda _0 \ne 0$. From (\ref{eq50}) it follows that the Lagrange multipliers $\lambda_1$ and $\lambda_2$ satisfy the equation
\begin{equation}
{\partial \lambda _2 \over \partial t} + \sum _{i=1} ^s \left ({\partial \lambda _2 \over \partial q_i}{\partial H \over \partial p_i} - {\partial \lambda _2 \over \partial p_i}{\partial H \over \partial q_i}\right ) =  \lambda _1 . \label{eq52}
\end{equation}

In this variational problem, the function $D(q,p,t | (q_0,p_0)_\omega ,t_0 )$ that renders $J[D]$ in (\ref{eq41}) stationary subject to constraints (\ref{eq43}) and (\ref{eq44}), is not required to take on prescribed values on the boundary of integration region $M \times (t_0,t_a)$. Therefore, it is necessary, that in addition to satisfying the Euler equation (\ref{eq50}), it also satisfies  the Euler boundary condition on the boundary of $M \times (t_0,t_a)$, ref. \cite{wan}. For all points on the portion of the boundary of $M \times (t_0,t_a)$ where time $t = t_0$ or $t = t_a$, the Euler boundary condition gives independently:
\begin{equation}
\left [\frac{\partial K}{\partial (\partial _t D)} - \lambda _2F\right ] _{t = t_0, t_a} = - \left [ \log D + \lambda _2\right ]_{t = t_0, t_a} F = 0 .  \label{eq53}
\end{equation}
For all points on the portion of the boundary of $M \times (t_0,t_a)$ where time $t$ is in the interval $t_0 < t < t_a$, the Euler boundary condition gives:
\begin{equation}
F \left [\lambda _2 {\bf v} \cdot {\bf n}\right ]_{\ \mathrm {at \ the \ boundary \ of} \ M}  = 0 . \label{eq55}
\end{equation}
In (\ref{eq55}), ${\bf v} \cdot {\bf n}$ is a scalar product of the velocity field ${\bf v} (q,p)$ in $\Gamma $ (defined in Sect. \ref{secHD}) and the unit normal ${\bf n}$ of the boundary surface of invariant set $M$, taken at the surface. Equation (\ref{eq55}) is satisfied naturally due to Hamiltonian motion, since the set $M$ is invariant by definition, and therefore  ${\bf v} \cdot {\bf n} = 0$ for all points on the boundary surface of $M$. This is a consequence of the fact that phase space paths do not cross over the boundary surface of the invariant set $M$.   
 
Functions $D(q,p,t | (q_0,p_0)_\omega ,t_0 )$ that render $J[D]$ in (\ref{eq41}) stationary subject to the constraints (\ref{eq43}) and (\ref{eq44}) are determined from the constraints and the boundary condition given by (\ref{eq53}). From (\ref{eq53}) one obtains the form of $D(q,p,t | (q_0,p_0)_\omega ,t_0 )$ at times $t_0$ and $t_a$, 
\begin{equation}
\left. D(q,p,t | (q_0,p_0)_\omega ,t_0 ) \right |_{t = t_0, t_a} = \left. \exp \left [ - \lambda _2 ((q_0,p_0)_\omega ,t_0; q,p,t)\right ] \right |_{t = t_0, t_a} . \label{eq56}
\end{equation}
Since it is only required that $t_a \geq t_0$, time $t_a$ is arbitrary in other respects. The boundary condition (\ref{eq53}) then holds for any time $t \geq t_0 $: 
\begin{equation}
D(q,p,t | (q_0,p_0)_\omega ,t_0 )  = \exp \left [ - \lambda _2 ((q_0,p_0)_\omega , t_0 ; q,p ,t ) \right ] . \label{eq60}
\end{equation}
From the constraint (\ref{eq44}), using (\ref{eq60}), one immediately obtains an equation for the Lagrange multiplier $\lambda _2 ((q_0,p_0)_\omega ,t_0; q,p,t)$:
\begin{equation}
{\partial \lambda _2 \over \partial t} + \sum _{i=1} ^s \left ({\partial \lambda _2 \over \partial q_i}{\partial H \over \partial p_i} - {\partial \lambda _2 \over \partial p_i}{\partial H \over \partial q_i}\right ) =  0 . \label{eq61} 
\end{equation}
By comparison of (\ref{eq52}) with (\ref{eq61}), it follows that for all $t \geq t_0 $,
\begin{equation}
\lambda _1 ((q_0,p_0)_\omega ,t_0; t) = 0 . \label{eq62}
\end{equation}
As explained in Sect. \ref{secIE}, for any physically well defined conditional probability distribution $D(q,p,t | (q_0,p_0)_\omega ,t_0 )$ (in the sense of (\ref{eq21}) and (\ref{eq23})), the upper bound on $S_{I}^{DF}(t,t_0)$, given by (\ref{eq33}), is not attained in maximization.

The conclusions that follow from the interpretation of (\ref{eq33}) and the property of $S_{I}^{DF}(t,t_0)$ as a measure of uncertainty explained in Sect. \ref{secIE}, are considered now in the second approach. This is done by replacing the strict equality constraint (\ref{eq44}) by the constraint which is of isoperimetric form,  
\begin{equation}
\varphi _2 ((q_0,p_0)_\omega ,t_0; t, D) =  \int_M \left [{\partial D \over \partial t} + \sum _{i=1} ^s \left ({\partial D \over \partial q_i}{\partial H \over \partial p_i} - {\partial D \over \partial p_i}{\partial H \over \partial q_i}\right )\right ]F \ d\Gamma = 0 . \label{eq63}
\end{equation}
The functional (\ref{eq46}) is then replaced with the functional
\begin{equation}
C_2[D, \lambda _2] = \int_{S_0(M)} \int _{t_0}^{t_a} \lambda _2\varphi _2 \ dt dS_0. \label{eq64}
\end{equation}  
Lagrange multiplier $\lambda _2 \equiv \lambda _2((q_0,p_0)_\omega ,t_0; t)$ is now a function defined in the integration region in the integral (\ref{eq64}). Information that the set $M$ of possible microstates is invariant to Hamiltonian motion is contained in the constraint (\ref{eq63}). Analogy is not complete, because a much larger class of functions satisfies the constraint (\ref{eq63}), including all functions that in addition, satisfy also the constraint (\ref{eq44}). This fact allows for maximization of the conditional information entropy $S_{I}^{DF}(t_a,t_0)$ in (\ref{eq40}), subject to constraints (\ref{eq43}) and (\ref{eq63}), even if $D(q,p,t | (q_0,p_0)_\omega ,t_0 )$  is prescribed at initial time $t_0$. The prescribed $D(q,p,t | (q_0,p_0)_\omega ,t_0 )$ at initial time $t_0$ must be physically well defined in the sense of (\ref{eq21}) and (\ref{eq23}).  In this variational problem, function $D(q,p,t | (q_0,p_0)_\omega ,t_0 )$  is not required to take on prescribed values on the remaining portion of the boundary of integration region $M \times (t_0,t_a)$ in (\ref{eq40}).

For a function $D(q,p,t | (q_0,p_0)_\omega ,t_0 )$ to maximize $S_{I}^{DF}(t_a,t_0)$ subject to constraints (\ref{eq43}) and (\ref{eq63}), it is necessary that it satisfies the Euler equation:   
\begin{eqnarray}
\lambda _0 \left \{\frac{\partial K}{\partial D} - \frac{d}{dt}\left (\frac{\partial K}{\partial (\partial _t D)}\right ) - \sum_ {i = 1}^s \left [\frac{d}{dq_i}\left (\frac{\partial K}{\partial (\partial _{q_i} D)}\right ) + \frac{d}{dp_i}\left (\frac{\partial K}{\partial (\partial _{p_i} D)}\right ) \right ]\right \} - \lambda _1F + {\partial \lambda _2 \over \partial t} F   = 0 . \label{eq65}
\end{eqnarray}
Another necessary condition for a maximum, in addition to (\ref{eq65}), exists if function $D(q,p,t | (q_0,p_0)_\omega ,t_0 )$ is not required to take on prescribed values on a portion of the boundary of $M \times (t_0,t_a)$: then, it is necessary that $D(q,p,t | (q_0,p_0)_\omega ,t_0 )$ satisfies the Euler boundary condition on the portion of the boundary of $M \times (t_0,t_a)$ where its values are not prescribed, ref. \cite{wan}. In accordance with this, for all points on the portion of the boundary of $M \times (t_0,t_a)$ where $t = t_a$, the Euler boundary condition gives: 
\begin{equation}
\left [\frac{\partial K}{\partial (\partial _t D)} - \lambda _2F\right ] _{t = t_a}  = - \left [ \log D + \lambda _2\right ]_{t = t_a} F = 0 . \label{eq68}
\end{equation}
The Euler boundary condition is satisfied naturally for all points on the portion of the boundary of $M \times (t_0,t_a)$ where time $t$ is in the interval $t_0 < t < t_a$. The set $M$ is invariant to Hamiltonian motion, and equation analogous to (\ref{eq55}) is also satisfied here naturally due to Hamiltonian motion.   

In analogous manner leading to (\ref{eq52}) in the first approach, the Euler equation (\ref{eq65}) now leads to the equation for the Lagrange multipliers $\lambda _1 ((q_0,p_0)_\omega ,t_0; t)$ and $\lambda _2 ((q_0,p_0)_\omega ,t_0; t)$:
\begin{equation}
{\partial \lambda _2 \over \partial t} =  \lambda _1  . \label{eq66}
\end{equation}
The form of the MaxEnt conditional probability distribution at time $t_a$ follows from (\ref{eq68}):  
\begin{equation}
D(q,p,t_a | (q_0,p_0)_\omega ,t_0 )  = \exp \left [ - \lambda _2 ((q_0,p_0)_\omega ,t_0; t_a) \right ] . \label{eq70}
\end{equation}
For a well defined conditional probability distribution at initial time $t_0$, there is an entire class of equally probable solutions $D(q,p,t | (q_0,p_0)_\omega ,t_0 )$ obtained by MaxEnt algorithm, which all satisfy the macroscopic constraint (\ref{eq63}). At time $t_a$, all functions in this class of MaxEnt solutions are equal and given by (\ref{eq70}). With the exception of times $t_0$ and $t_a$, the conditional probability distribution $D(q,p,t | (q_0,p_0)_\omega ,t_0 )$ obtained by MaxEnt algorithm is not uniquely determined in the interval \linebreak $t_0 < t < t_a$. This is a consequence of the fact that the macroscopic constraint (\ref{eq63}) does not determine the time evolution of $D(q,p,t | (q_0,p_0)_\omega ,t_0 )$ uniquely, in the way that the strict microscopic constraint (\ref{eq44}) does. However, MaxEnt solutions still predict only time evolutions entirely within the invariant set $M$, due to (\ref{eq55}). This property follows from the constraint (\ref{eq63}), and takes into account the information about the constants of motion that determine the invariant set $M$, and in this way, about related conservation laws.

From the normalization (\ref{eq37}) of the conditional probability distribution, given at time $t_a$ by (\ref{eq70}), one obtains the relation:  
\begin{equation}
W(M) \exp \left [ - \lambda _2 ((q_0,p_0)_\omega ,t_0; t_a) \right ] = 1 , \label{eq70a} 
\end{equation}
where $W(M)$ is the measure, i.e., phase space volume of the invariant set $M$. Equation (\ref{eq70a}) implies that the Lagrange multiplier $\lambda _2 ((q_0,p_0)_\omega ,t_0; t)$ at time $t = t_a$ is independent of the variables $(q_0,p_0)_\omega $:
\begin{equation}
\lambda _2 ((q_0,p_0)_\omega ,t_0; t_a) = \lambda _2 (t_a) . \label{eq71}
\end{equation}
Microstate probability distribution $f(q,p,t)$ at time  $t = t_a$ is then calculated by using: (\ref{eq17}) and (\ref{eq23}), the MaxEnt conditional probability distribution $D(q,p,t | (q_0,p_0)_\omega ,t_0 )$ at time  $t = t_a$ given by (\ref{eq70}) and (\ref{eq71}), and the path probability distribution $F((q_0,p_0)_\omega ,t_0 )$ at initial time $t_0$:
\begin{equation}
f(q,p,t_a) = \exp \left [ - \lambda _2 (t_a) \right ] . \label{eq72}
\end{equation}
It follows from (\ref{eq70}--\ref{eq72}) that at time  $t_a$, the MaxEnt conditional probability distribution and the corresponding microstate probability distribution  are equal,
\begin{equation}
D(q,p,t_a | (q_0,p_0)_\omega ,t_0 ) = f(q,p,t_a) = \exp \left [ - \lambda _2 (t_a) \right ] = \frac{1}{W(M)} . \label{eq73}
\end{equation}
From (\ref{eq28}), (\ref{eq31}) and (\ref{eq73}), one obtains the values of information entropies $S_{I}^{DF}(t,t_0)$ and $S_I ^f (t)$  at time $t_a$,
\begin{equation}
S_I ^f (t_a) = S_{I}^{DF}(t_a) = \log W(M) . \label{eq74} 
\end{equation}
Equations (\ref{eq73}) and (\ref{eq74}) are possible only in case of statistical independence, i.e.,  the complete loss of correlation between the phase space paths at time $t_0$, and the microstates at time $t_a$. In general, property of macroscopic systems is that they appear to randomize themselves between observations, provided that the observations follow each other by a time interval longer then a certain characteristic time $\tau $ called the relaxation time \cite{kittel}. In the interpretation given here, relaxation time $\tau $ for a closed Hamiltonian system represents a characteristic time required for the described loss of correlation between the initial phase space paths and final microstates. Furthermore, $\tau $ also represents a time interval during which predictions, based on incomplete information about microscopic dynamics, become uncertain to a maximum extent compatible with data. This uncertainty is related to loss of information about the state of the system. 

This interpretation is reflected in the role of the Lagrange multipliers $\lambda _1 ((q_0,p_0)_\omega ,t_0; t)$ and $\lambda _2 ((q_0,p_0)_\omega ,t_0; t)$. They are required to satisfy (\ref{eq66}), and by integrating it one obtains the following relation, 
\begin{equation}
 \lambda _2 ((q_0,p_0)_\omega ,t_0; t) = \int_{t_0}^{t} \lambda _1 ((q_0,p_0)_\omega ,t_0; t^{\prime})dt^{\prime} + \lambda _2 ((q_0,p_0)_\omega ,t_0; t_0) , \label{eq75}
\end{equation} 
for all $t$ in the interval $t_0 \leq  t \leq  t_a$. By using (\ref{eq75}), with (\ref{eq70a}), (\ref{eq71}) and (\ref{eq74}), one obtains
\begin{equation}
 S_I ^f (t_a) = S_{I}^{DF}(t_a,t_0) = \log W(M) = \int_{t_0}^{t_a} \lambda _1 ((q_0,p_0)_\omega ,t_0; t)dt + \lambda _2 ((q_0,p_0)_\omega ,t_0; t_0) . \label{eq77}
\end{equation} 
It is clear, from relations (\ref{eq71}), (\ref{eq75}) and (\ref{eq77}), that at time $t_a$ the Lagrange multiplier $\lambda _2 ((q_0,p_0)_\omega ,t_0; t_a ) \equiv \lambda _2 (t_a)$ is determined by the measure $W(M)$ of the invariant set $M$ of possible microstates, i.e., the volume of accessible phase space. The subsequent application of MaxEnt algorithm of the described type for a closed system with Hamiltonian dynamics, without the introduction of additional constraints, results in the increase of  $W(M)$. From (\ref{eq71}), (\ref{eq75}) and (\ref{eq77}) it is then deduced that $\lambda _2(t_a) \geq  \lambda _2(t_0)$.

Information about the structure of possible microstates restricts the corresponding set, and therefore sets an upper bound on the volume of accessible phase space. The values of $S_{I}^{DF}(t_a,t_0)$ and $S_I ^f (t_a)$ at time $t_a$, given in (\ref{eq77}), are equal to the maximum value of the {\it Boltzmann-Gibbs entropy}, compatible with this information. The Lagrange multiplier $\lambda _1 ((q_0,p_0)_\omega ,t_0; t)$, integrated in (\ref{eq77}) over time $t_0 \leq t \leq t_a$, is then determined by the rate at which the maximum Boltzmann-Gibbs entropy is attained in reproducible time evolution. The integral in (\ref{eq77}), and the quantity $\lambda _1 ((q_0,p_0)_\omega ,t_0; t)$, can be identified with the change in entropy, and the {\it entropy production} for a closed Hamiltonian system, respectively. If information about microscopic dynamics of closed Hamiltonian system is considered complete, whether entropy production can be defined without recourse to coarse graining procedures, or macroscopic, phenomenological approaches, remains an open question. In general, information is discarded in all such models, at some stage, in order to match with what is observed in nature.

\section{Conclusion and related issues} 
If we consider the possibility that our information about microscopic dynamics is incomplete, reproducibility and information loss become a part of description of the macroscopic time evolution. This approach then leads to a simple definition for entropy production. The idea that irreversibility is related to a gradual loss of information has been developed by Jaynes in the density-matrix formalism \cite{jaynes2}. Recently, Duncan and Semura \cite{dunsem1,dunsem2} suggested the notion that information is really lost at a fundamental level. The interplay of quantum decoherence and dynamics is considered as one of possible reasons behind the second law of thermodynamics, with entropy production caused by information leaking into the environment \cite{zurek2,zurek3}. In our classical approach, information loss is related to incomplete information about microscopic dynamics. If one considers this idea carefully, even in this simple model, incompleteness of information must be taken into account in some unbiased way.   

The issues related to incomplete information can be discussed in an objective manner. Analytical principles for such purposes, i.e. for a separation of the subjective and objective aspects of the theoretical formalism, are found in probability theory. Philosophy of this approach is based on the interpretation of probability theory as a natural extension of deductive logic. Such generalization has been developed in axiomatic way by Jaynes in his treatise on probability theory \cite{jaynes9}. It was intended as a tool for plausible inference in situations of incomplete information. The standard axiomatic probability theory is derived from this generalized theory, suggesting in itself that the generalized theory is a proper tool for incorporating new information in our probability distributions. Probability distributions are interpreted in that sense as carriers of incomplete information. This approach is perhaps best understood from descriptions given by Jaynes \cite{jaynes9}: ``\dots  probability theory as a generalized logic of plausible inference which should apply, in principle, to any situation where we do not have enough information to permit deductive reasoning." We quote also the following lines from \cite{jaynes9}, which we think are important for the discussion in the next paragraph: ``But this is equally true of abstract mathematical systems; when a proposition is undecidable in such a system, that means only that its axioms do not provide enough information to decide it. But new axioms, external to the original set, might supply the missing information and make the proposition decidable after all." We can conclude that when probabilities are interpreted in a related way as a property of our state of knowledge, and applied supplemented with MaxEnt algorithm, that the mathematical description of irreversible behavior fits naturally within the concepts of Shannon's information theory \cite{shannon}.

Another objective aspect of the problem mentioned above is related to the issues that were raised in very interesting and speculative way by Zwick. In his paper \cite{zwick} on the measurement problem in quantum mechanics, the difficulty of describing it at the level of quantum dynamics is compared, and found to be similar with the incompleteness of certain axiomatic systems in mathematics, discovered and elaborated by G\"odel and others \cite{godel,nagel}. According to Zwick, the extensive parallelism between the physical and mathematical cases suggest the possibility that the measurement process is self-referential as was G\"odel's special formula, and that measurement may be undecidable within the dynamics (formalism of the time-dependent Schr\"odinger equation), and occurring only at a meta-level of the formalism. In such line of thinking, physical theory would have then at least two levels; the measurement process would be described at a meta-level, but undecidable on the base level which is described by the dynamical law. At the same time, the base level is inherently incomplete and no contradiction is generated. The dynamical law and all the processes described by it are reversible, but the measurement process is irreversible; in this way irreversibility would be present in a description of the measurement process within such two-level theory. Zwick quotes similar suggestions by Pattee \cite{pattee} about the necessity of two levels of structure and description for any prediction and control (i.e. measurement) process. Questions that are raised about irreversible behavior of systems governed deterministically by the time-symmetric equations of motion, would then appear paradoxical only in the context of single-level theory \cite{pattee}. 

Without involving us more deeply in these issues, we note that in our application of MaxEnt to the problem of prediction of time evolution of closed systems with Hamiltonian dynamics, certain features of two level theory can be clearly recognized. In this simple model they appear only as a result of  our recognition of incompleteness of our own information about microscopic dynamics. From pragmatic viewpoint this allows us to discuss further on the related issues about the interplay between our knowledge and measurement constraints on the system and its ``actual" dynamics. 

\begin{acknowledgments}
Authors whish to thank the anonymous reviewer for insightful suggestions that significantly improved the submitted manuscript. The present work was supported by Croatian MZOS project no. 177-1770495-0476.
\end{acknowledgments}

\end{document}